\documentstyle[11pt]{article}

\begin{document}

\bibliographystyle{unsrt}

\title{Strong enhancement of current, efficiency and mass separation
in Brownian motors driven by non Gaussian noises}

\author{Sebastian Bouzat \thanks{E-mail:
bouzat@cab.cnea.gov.ar} and Horacio S. Wio \thanks{E-mail
wio@cab.cnea.gov.ar} \\ Grupo de F\'{\i}sica Estad\'{\i}stica
\thanks{http://www.cab.cnea.gov.ar/Cab/invbasica/FisEstad/estadis.htm},
\\
Centro At\'omico Bariloche (CNEA) and Instituto Balseiro (UNC)
\\
8400 San Carlos de Bariloche, R\'{\i}o Negro, {\bf Argentina}}

\date{December 14th, 2001}
\maketitle

\begin{abstract}
We study a Brownian motor driven by a colored non Gaussian noise
source with a $q$-dependent probability distribution, where $q$ is
a parameter indicating the departure from Gaussianity. For $q=1$
the noise is Gaussian (Ornstein--Uhlenbeck), while, for $q>1$, the
probability distribution falls like a power law. In the latter
case, we find a marked enhancement of both the current and the
efficiency of the Brownian motor in the overdamped regime. We also
analyze the case with inertia and show that, again for $q > 1$, a
remarkable increase of the ratchet's mass separation capability is
obtained.
\end{abstract}

\newpage

The study of noise induced transport by "ratchets" has attracted
an increasing number of researchers that have produced a vast
literature \cite{inic,magnasco,tilting,pulsating,invers}. Among
other aspects, the motivation of these studies has been prompted
by both their possible biological interest \cite{inic} as well as
their potential technological applications. Since the initial
works, besides the built-in ratchet-like bias and correlated
fluctuations (see for instance \cite{magnasco}), different aspects
have been studied like tilting \cite{tilting} and pulsating
\cite{pulsating} potentials, velocity inversions \cite{invers},
etc. There are some reviews where it is possible to grasp the
state of the art \cite{review1,review2}.

So far, almost all studies have used Gaussian noises, with few
exceptions that mainly exploited dichotomic processes
\cite{review2}. Here we analyze the effect of a particular class
of colored non Gaussian noise on the transport properties of
Brownian motors. Such a noise source is based on the so called
Tsallis statistics \cite{tsallis} with a probability distribution
that depends on $q$,  a parameter indicating the departure from
the Gaussian ($q = 1$) behavior. Some of the motivations for
studying the effect of non Gaussian noises are, in addition to the
intrinsic interest within the realm of noise induced phenomena,
different experimental data indicating that for some biologically
motivated systems fluctuations can have a non Gaussian character.
An example are current measurements through voltage-sensitive ion
channels in a cell membrane or experiments on the sensory system
of rat skin \cite{nature}. It is also worth remarking here that
recent detailed studies on the source of fluctuations in some
biological systems \cite{caltech} clearly indicate that noise
sources in general could be non Gaussian and their distribution
bounded. Here, again as a consequence of the non Gaussian
character of the driven noise, we find a remarkable increase of
the current together with an enhancement of the motor efficiency
that, additionally, shows an optimum for a given degree of non
Gaussianity. Moreover, when the inertia is taken into account we
find that, when departing from the Gaussian case, there is a
remarkable increment in the mass separation efficiency of these
devices.

We start considering the general system
\begin{equation}
\label{xpto} m\frac{d^2 x}{dt^2}=-\gamma\frac{dx}{dt}
-V'(x)-F+\xi(t)+\eta(t),
\end{equation}
where $m$ is the mass of the particle, $\gamma$ the friction
constant, $V(x)$ the ratchet potential, $F$ is a constant ``load''
force, and $\xi(t)$ the thermal noise satisfying $\langle \xi(t)
\xi(t') \rangle= 2 \gamma KT \delta(t-t')$. Finally, $\eta(t)$ is
the time correlated forcing (with zero mean) that keeps the system
out of thermal equilibrium allowing the rectification of the
motion.  For this type of ratchet model several different kinds of
time correlated forcing have been considered in the literature
\cite{review1,review2}. We may distinguish the cases of
deterministic (such as the periodic forcing and the ``synthetic
noise'' discussed in the pioneering work of Magnasco
\cite{magnasco}) and of stochastic forcing. Among the stochastic
forcings the more usual noise sources are dichotomic and
Ornstein--Uhlenbeck (OU) noises \cite{review2}. Here, as indicated
above, we will study the effects of a different family of
stochastic forcing having non Gaussian statistics. The main
characteristic introduced by the non Gaussian form of the forcing
we consider here is the appearance of arbitrary strong ``kicks''
with relatively high probability when compared, for example, with
the OU Gaussian process. As we shall see, in a general situation
(without fine tuning of the parameters), this leads to the above
indicated remarkable effects.

We will consider the dynamics of $\eta(t)$ as described by the
Langevin equation \cite{qRE1,qRE2}
\begin{equation}
\label{etapto} \frac{d\eta}{dt}=-\frac{1}{\tau}\frac{d}{d \eta}
V_q(\eta) + \frac{1}{\tau} \zeta(t),
\end{equation}
with  $\langle\zeta(t)\zeta(t)'\rangle=2D\delta(t-t')$ and
\begin{equation}
V_q(\eta)=\frac{D}{\tau(q-1)}\ln[1+\frac{\tau}{D}(q-1)\frac{\eta^2}{2}].
\end{equation}
Previous studies of such processes in connection with
stochastic resonance problems \cite{qRE1,qRE2} and dynamical
trapping \cite{qruido} have shown that the non Gaussianity of
the noise leads to interesting effects. For $q=1$,
the process $\eta$ is the (Gaussian) Ornstein--Uhlenbeck
one (with correlation time equal to $\tau$), while for $q\neq 1$
it is a non Gaussian process. For $q<1$ the probability
distribution has a cut--off at
$\omega=[(1-q)\tau/(2D)]^{-\frac{1}{2}}$, and is given by:
\begin{equation}
P_{q<1}(\eta)=\left\{ \begin{array}{cc} \frac{1}{Z_q}
[1-(\frac{\eta}{\omega})^2]^{\frac{1}{1-q}}
&{\rm if} |\eta|<\omega. \\
0&{\rm otherwise} \end{array} \right.
\end{equation}
where $Z_q$ is the normalization constant. For $1<q<3$, the
probability distribution (for $-\infty<\eta<\infty$) is
\begin{equation}
P_{q>1}(\eta)=\frac{1}{Z_q}\left[1+\frac{\tau}{2
D}(q-1)\eta^2\right]^\frac{1}{1-q}.
\end{equation}
While keeping $D$ constant, the width or dispersion of the
distribution increases with $q$. This means that, the higher the
$q$, the stronger the ``kicks'' that the particle will receive.
The second moment of the distribution (or intensity of the non
Gaussian noise, that we call $D_{ng}$) diverges for $q\ge 5/3
\simeq 1.66$ while, for $q\le5/3$, is given by
\begin{equation}
\label{Dng} D_{ng} \equiv \langle \eta\rangle= \frac{2 D}{\tau
(5-3q)}.
\end{equation}
Here, and in what follows, we shall keep $q<5/3$.

For the ratchet potential we will first consider the same form as
in \cite{magnasco} (with period $2 \pi$)
\begin{equation}
V(x)=- \int dx (\exp[\alpha \cos(x)]/J_0(i\alpha)-1),
\end{equation}
with $\alpha=16$. The integrand is the ratchet force (-V'(x))
appearing in Eq.(\ref{xpto}).

Firstly, we will analyze the overdamped regime setting $m=0$ and
$\gamma=1$. We are interested on analyzing the dependence of the
mean current $J=\langle \frac{dx}{dt} \rangle$ and the efficiency
$\varepsilon$  on the different parameters. In particular, their
dependence on $q$, the parameter indicating the degree of non
Gaussianity of the noise distribution. The efficiency of the
ratchet system is defined as the ratio of the work (per unit time)
done by the particle ``against'' the load force $F$, into the mean
power injected to the system through the external forcing $\eta$
\cite{Energ1}.
\begin{equation}
\varepsilon =\frac{\frac{1}{T_f}\int_{x=x(0)}^{x=x(T_f)} F
dx(t)}{\frac{1}{T_f}\int_{x=x(0)}^{x=x(T_f)} \eta(t) dx(t)}
\end{equation}
For the numerator we get $F\langle \frac{dx}{dt} \rangle = F J$,
while, for the denominator
\begin{equation}
\frac{1}{T_f}\int_0^{T_f} \eta (t) \frac{dx}{dt} dt =
\frac{1}{\gamma T_f} \int_0^{T_f}\eta(t)^2 dt =\frac{2 D}{\gamma
T_f \tau (5-3q)}.
\end{equation}
Interesting and complete
discussions on the thermodynamics and energetics of ratchet
systems can be found in \cite{Seki}.

It is worth mentioning here that the ``effective Markovian
approximation'' introduced in \cite{qRE2} is not adequate for the
present case as it only properly works in the low $\tau$ limit.
Here, in the overdamped regime we are able to give an approximate
analytical solution for the problem, which is expected to be valid
in the large correlation time regime ($\frac{\tau}{D}
>>1$): we perform the adiabatic approximation of solving the
Fokker-Planck equation associated to Eq. (\ref{xpto}) assuming a
constant value of $\eta$ \cite{comFP}. This leads us to obtain a
value for the current $J(\eta )$ which depends on $\eta$  and, in
order to get the final result, we should perform an average of
$J(\eta)$ over $\eta$ using the distribution $P_q(\eta)$ with the
desired value of $q$.

In Fig. 1, we show typical analytical results for the current and
the efficiency as functions of $q$ together with results coming
from numerical simulations (for the complete system given by
Eqs.(\ref{xpto}) and (\ref{etapto})). Calculations have been done
in a region of parameters similar to the one studied in
\cite{magnasco} but considering (apart from the difference
provided by the non Gaussian noise) a non--zero load force that
leads to a non--vanishing efficiency. As can be seen, although
there is not a quantitative agreement between theory and
simulations, the adiabatic approximation predicts qualitatively
very well the behavior of $J$ (and $\varepsilon$) as $q$ is
varied. As shown in the figure, the current grows monotonously
with $q$ (at least for $q<5/3$) while there is an optimal value of
$q$ ($ > 1$) which gives the maximum efficiency. This fact is
interpreted as follows: when $q$ is increased, the width of the
$P_q(\eta)$ distribution grows and high values of the non Gaussian
noise become more frequent, this leads to an improvement of the
current. Although the mean value of $J$ increases monotonously
with $q$, the grow of the width of $P_q(\eta)$ leads to an
enhancement of the fluctuations around this mean value. This is
the origin of the efficiency's decay that occurs for high values
of $q$: in this region, in spite of having a large (positive) mean
value of the current, for a given realization of the process, the
transport of the particle towards the desired direction is far
from being assured.

In Fig. 2 we show results from simulations for $J$ and
$\varepsilon$ as functions of $q$ for different values of $D$, the
intensity of the white noise in Eq. (\ref{etapto}). The results
correspond to $KT=0$, hence, the only noise present in the system
is the non Gaussian one. On the curve corresponding to the results
for $J$ (Fig. 2.a.), we indicate with error bars the dispersion of
the results. The huge growth of the dispersion occurring for $q >
1.3$, as well as the decay of the efficiency for the same values
of $q$, is apparent.

It is worth recalling here that $q=1$ corresponds to the Gaussian
OU noise (analyzed, for example in \cite{magnasco} and
\cite{OUrat}). Hence, our results show that the transport
mechanism becomes more efficient when the stochastic forcing has a
non Gaussian distribution with $q>1$.

In the previous calculations, we have analyzed the values of $J$
and $\varepsilon$ as functions of $q$ for fixed values of $D$.
However, there are situations where we should consider that the
non Gaussian noise source is the ``primary'' source, for instance
see \cite{caltech} for biologically motivated problems. Besides
this, there could also be situations of clear technological
interest. For these reasons we wondered which would be the
dependence of $J$ and $\varepsilon$ on $q$ when we fix the
intensity of the non Gaussian noise $D_{ng}$ defined in
Eq.(\ref{Dng}). In such case, the energy per unit time supplied to
the ratchet is independent of $q$. Our results (not shown)
indicate that, at variance to what occurs in calculations for
fixed $D$, for constant $D_{ng}$, the optimum value of $q$ which
maximizes the efficiency also maximizes the mean current $J$,
which decays for larger values of $q$. This result is easy to
understand by observing that the width of the $P_q(\eta)$
distribution decreases with $q$ when $D_{ng}$ is kept constant.

Now we turn to study the $m \neq 0$ case, that is, the situations
in which the inertia effects are relevant. In Fig. 3 we show the
dependence of the current $J$ on the mass $m$ for different values
of $q$. The results are from simulations for zero temperature and
without load force. It can be seen that, as $m$ is increased from
$0$, the inertial effects initially contribute to increase the
current, until an optimal value of $m$ is reached. As it is
expected, for high values of $m$, the motion of the particle
becomes difficult and, for $m\to \infty$, the current vanishes.

An interesting effect appears for $q=1.3$: for a well defined
interval of the value of the mass (ranging approximately from
$m=100$ to $m=5000$), a negative current is observed. This can be
explained as a consequence of the high value of the mass, that
makes the inertial effects much more important than those of the
ratchet potential. Due to the high value of $m$, most of the time
the particle will be almost at rest around the minimum of the
ratchet potential. However, a very high value of $\eta$ may occur
(that may last for an interval of time of the order of $\tau$)
causing the particle to move a period of the potential to the
right or to the left. The obvious question is, what is the value
of the force $\eta$ needed in order to make the particle jump a
period of the potential to the left (right), during a time
$\tau$?. Assuming no friction (which is of no importance for high
values of $m$), a constant (average) value of the ratchet force
$V'$, and performing a classical calculation we get:
\begin{equation}
\label{clasapp} \eta=|V'|+\frac{2 m d}{\tau^2},
\end{equation}
where $d$ is the distance from one minimum of $V(x)$ to the next
left (right) maximum of $V(x)$. In our system, for a jump to the
left, we have $|V'| \sim 4.5$ and $d \sim 1.1$, while, for a jump
to the right, we have $|V'| \sim .96$ and $d \sim 5.2$.
Considering $\tau=100/(2 \pi)$, for $m=800$ (the value for which
in Fig. 3 the minimum of the current is obtained for $q=1.3$) we
find that the required value of $\eta$ for a jump to the left
(right) is $|\eta_l| \sim 6.9$ ($|\eta_r| \sim 32.4$). Hence, for
this value of the mass, the jumps to the left are more probable
than the jumps to the right, and a negative current is to be
expected. In contrast, for a value of the mass $m=100$
(approximately where the current change from being positive to
being negative) we find $|\eta_l| \sim 5.4$ and $|\eta_r| \sim
5.1$, i.e. both jumps are almost equiprobable. Note that, for
$m=800$, the large value of $m$ makes the second term in Eq.
(\ref{clasapp}) more important than the contribution of $|V'|$
(the inertial effect dominates over the ratchet force), hence, the
difference in the value of $\eta$ necessary for a jump to the left
or to the right comes, essentially, from the difference on the
distances ($d$) that the heavy particle should be displaced to
each side.

Until this point, the analysis makes no mention of the value of
$q$ of the noise distribution. The fact that the $(J<0)$--effect
appears for $q=1.3$ and not for $q=1$ is understood as a
consequence of the fact that, for $q=1$, the occurrence of a value
of noise $|\eta_L|$ is highly improbable in the regime where the
inertia dominates over the effect of the ratchet force. Hence, the
particle remains essentially motionless.

These results imply that separation of masses (particles with
different masses moving in opposite directions) occurs, and that
this happens in the absence of load force, due to the ``big help''
given by the non Gaussian noise. To our knowledge, separation of
masses by ratchets with zero load force have not being reported
before. However, the separation found here occurs for particles
with a ratio of masses of the order of $10$ or more (say $800/80$
in Fig. 3.), while it has been shown in \cite{Lind,Lind2} that,
with a load force and considering simply OU noise, particles of
much closer masses can be separated by ratchets.

Now we analyze in more detail the problem of mass separation, an
aspect that has been studied in some works
\cite{Lind,Landa,Lind2}. In view of the results discussed above,
it is reasonable to expect that non Gaussian noises may improve
the capability of mass separation in ratchets in more general
situations. Reference \cite{Lind} was one of the primary works
discussing mass separation by ratchets. There, the authors
analyzed a ratchet system like the one described by Eq.
(\ref{xpto}) considering OU noise as external forcing (in our case
it corresponds to $q=1$). They studied (both numerically and
analytically) the dynamics for different values of the correlation
time of the forcing $\tau$, finding that there is a region of
parameters where mass separation occurs. This means that the
direction of the current is found to be mass--dependent: the
``heavy'' species moves in the negative sense while the ``light''
one, do so in the positive sense.

Here, in order to compare results, we analyze the same system
studied in \cite{Lind,Lind2} but considering the non Gaussian
forcing described by Eq.(\ref{xpto}). Hence, we study the system
in Eq.\ref{xpto} with  $V(x)=-[\sin(2 \pi x)+.25 \sin(4 \pi x)]/(2
\pi)$ as the ratchet potential. We focus on the region of
parameters where, in \cite{Lind} (for $q=1$), separation of masses
was found. We fix $\gamma=2, KT=.1, \tau=.75$, and $D=.1875$ and
consider the values of the masses $m=m_1=0.5$ and $m=m_2=1.5$ as
in \cite{Lind}. Our main result here is that the separation of
masses is enhanced when a non--Gaussian noise with $q>1$ is
considered. In Fig. 4.a. we show $J$ as function of $q$ for
$m_1=0.5$ and $m_2=1.5$. It can be seen that there is an optimum
value of $q$ that maximizes the difference of currents. In that
figure, this value, which is close to $q=1.25$, is indicated with
a vertical double arrow. Another double arrow indicates the
separation of masses occurring for $q=1$ (Gaussian OU forcing).
The calculations are for a load force $F=0.25$. We have observed
that, when the value of the load force is varied, the difference
between the curves remain approximately constant but both are
shifted together to positive or negative values (depending on the
sign of the variation of the loading). By controlling this
parameter it is possible to achieve, for example, the situation
shown in Fig. 4.b., where, for the value of $q$ at which the
difference of currents is maximal, the heavy ``species'' remains
static on average (has $J=0$), while the light one has $J>0$. Also
it is possible to get the situation shown in Fig. 4.c, at which
the two species moves in the opposite direction with equal
absolute velocity.

Summarizing, we have systematically studied the effect of a
colored non Gaussian noise source on the transport properties of a
Brownian motor. What we have found is that a departure from
Gaussian behavior, given by a value of $q$ larger than 1, induces
a remarkable increase of the current together with an enhancement
of the motor efficiency. The latter shows, in addition, an optimum
value for a given degree of non Gaussianity. When inertia is taken
into account we also find a considerable increment in the mass
separation capability.

It is worth mentioning that in the studies of the influence of non
Gaussian noises in stochastic resonance and other related
phenomena, the system's response enhancement occurs for $q<1$, as
discussed in \cite{qRE1,qRE2,qruido}. The reason can be traced
back to the dependence of the mean-first-passage-time with $q$,
and the interplay between transition rates and modulation
frequency. In contrast, in the present work, and due to the
``favorable'' influence of high values of the noise for increasing
the transport effect, we observe that the relevant effect occurs
in the opposite case, that is for $q>1$.

These studies could be of interest for possible relation with
biologically motivated problems \cite{inic,nature,caltech,nature2}
as well as for potential technological applications, for instance
in "nanomechanics" \cite{review1,review2}. More specific studies
of aspects in one or other area will be the subject of further
work. \\ \\

{\bf Acknowledgements:} The authors thanks to V.Grunfeld for a
critical reading of the manuscript. Partial support from CONICET,
Argentina is acknowledged. HSW wants to thank to the IMEDEA and
Universitat de les Illes Balears, Palma de Mallorca, Spain, for
the kind hospitality extended to him.

\newpage

\begin{figure}[tbp]
\caption{Current (a) and efficiency (b) as functions of $q$. The
solid line corresponds to the analytical results in the adiabatic
approximation while the line with squares shows results from
simulations. All calculations are for
$m=0,\gamma=1,KT=0.5,F=0.1,D=1$ and $\tau=100/(2 \pi)$.}
\end{figure}

\begin{figure}[tbp]
\caption{Current (a) and efficiency (b) as functions of $q$.
Results from simulations at $KT=0$ for $D=1$ (circles), $D=10$
(squares), and $D=20$ (triangles). All calculations are for
$m=0,\gamma=1, F=0.1$ and $\tau=100/(2 \pi)$.}
\end{figure}

\begin{figure}[tbp]
\caption{Current as function of the mass for different values of
$q$: circles $q=1$ (Gaussian noise case), squares $q=1.3$.}
\end{figure}

\begin{figure}[tbp]
\caption{Separation of masses: results from simulations for the
current as a function of $q$ for particles of masses $m=0.5$
(hollow circles) and $m=1.5$ (solid squares). Calculations for
three different values of the load force: $F=.025$ (a), $F=0.02$
(b) and $F=0.03$ (c).}
\end{figure}


\begin{thebibliography}{99}
\bibitem{comFP} In doing this, we use a piecewice linear
approximation for the ratchet potential which is
$V(x)=2.5-(5/1.1)x$ for $x<1.1$ and $V(x)=-2.5+[5/(2 \pi
-1.1)](x-1.1)$ for $1.1<x<2\pi$. We solve the Fokker--Planck
equation for $0\le x \le 2 \pi$ with periodic boundary condition
asking for continuity of the distribution and the current at
$x=1.1$.
\end{thebibliography}
\end{document}